\documentstyle[amsmath,amssymb]{article}

\def\be{\begin{eqnarray}}
\def\ee{\end{eqnarray}}
\def\nn{\nonumber}

\hoffset= - 1.0in         

\begin{document}

\hfill ITEP/TH-14/07

\bigskip

\centerline{\Large{
Higher Nilpotent Analogues of $A_\infty$-Structure
}}

\bigskip

\centerline{\it V.Dolotin, A.Morozov and Sh.Shakirov}

\bigskip

\centerline{ITEP, Moscow, Russia}

\bigskip

\centerline{ABSTRACT}

\bigskip

Higher nilpotent analogues of the $A-\infty$-structure,
$\big(D_n\big)^n =
\left(\sum_{i=1}^\infty m_n^{(i)}\right)^{\circ n} = 0$,
are explicitly defined on arbitrary simplicial complexes,
generalizing explicit construction of arXiv:0704.2609
for $n=2$.
These structures are associated with the higher nilpotent
differential $m_n^{(1)} = d_n$, satisfying $d_n^n =0$,
which is naturally defined on triangulated manifolds
(tetrahedral lattices). The deformation
$D_n = (I + \epsilon_n) d_n (I + \epsilon_n)^{-1}$ is defined with the help of
the $n$-versions of exterior product $\wedge_n$
and the $K_n$-operator.

\bigskip

\bigskip

\section{Introduction}

$A_\infty$-structure \cite{Ainf} is a sequence of operations
\be
m^{(i)}:\ V^{\times i} \rightarrow V, \ \ \ \
i=1,2,\ldots,\infty
\label{mop}
\ee
on a graded vector space $V$, satisfying the
nilpotency relation:
\be
\left(\bigoplus_{i=1}^\infty m^{(i)}\right)
\!\!\!\!\!\!\!\!\!\!\!\!\!\!\!
{\phantom{5^{5^{5^{5^5}}}}}^{\circ 2} = 0,
\label{A2}
\ee
which is actually an infinite set of relations
\be
\sum_{j=1}^k m^{(j)}\circ m^{(k-j)} = 0
\label{A2comp}
\ee
on $V^{\otimes k}$ for all integer $k\geq 2$.
Composition of operations is denoted through $\circ$,
in what follows it will often be omitted and we assume
that it should not cause any confusion.
The $A_\infty$ structure is a natural deformation of
a nilpotent BRST operator \cite{BRST} $m^{(1)}=Q=d$,
which satisfies $Q^2 = d^2 = 0$,
and plays an important role in BV theory \cite{BV}
of topological field theories \cite{tot}.

$A_\infty$-structure also arises as a natural deformation
of de Rham $d-\wedge$ structure \cite{didR,DMS}
from continuous to discrete manifolds (tetrahedral lattices)
and to simplicial complexes \cite{sico}.
Explicit construction of this deformation is recently
given in \cite{DMS} and it can be immediately generalized
-- in the spirit of non-linear-algebra paradigm \cite{nolal} --
to higher nilpotent analogue of the $A_\infty$-structure:
\be
\big(D_n\big)^{\circ n} =
\left(\bigoplus_{i=1}^\infty m_n^{(i)}\right)
\!\!\!\!\!\!\!\!\!\!\!\!\!\!\!
{\phantom{5^{5^{5^{5^5}}}}}^{\circ n} = 0.
\label{An}
\ee
The point is that on simplicial complexes the discrete
differential $d=d_2$ is in no way distinguished:
it has immediate generalizations to higher nilpotent
$d_n$, which satisfies
\be
\big(d_n\big)^{\circ n} = 0,
\ee
it is enough to write $(d_n f)_{ij} = f_i + \omega_n f_j$
with an $n$-th root of unity
$\omega_n = \exp \frac{2\pi ik}{n}$
instead of the naive $(df)_{ij} = f_i - f_j$
for a link $(ij)$ between the two adjacent vertices
$i$ and $j$ of the complex -- all other elements of the
DMS-construction \cite{DMS} will immediately follow.
We call (\ref{An}) the $A^n_\infty$-structure, keeping
the usual notation $A_\infty = A^2_\infty$ for (\ref{A2}).
Of course, this generalization is essentially discrete
and does not have immediate continuum limit,
however, this is hardly a serious drawback
for topological considerations.
Moreover, consideration of higher nilpotents opens
essentially new possibilities and can strongly enhance
the power of (generalized) de Rham calculus and
related quantum-field-theory methods in topological studies.
Such applications, however, are left beyond the scope of
this paper.

We begin in section \ref{A2str}
from a brief remind of the DMS construction \cite{DMS}
of the ordinary (quadratic nilpotent) $A_\infty$-structure
on arbitrary simplicial complex $M$.
Its straightforward generalization to arbitrary $n>2$ is
described in the next section \ref{Anstr}.

\section{$A_\infty$ structure on simplicial complex
\label{A2str}}

The DMS construction involves the following ingredients:

$\bullet$
Simplicial complex $M$ with some ordering (enumeration)
of vertices. Ordered complex is denoted $\bar M$.
Ordering is important for definition of sign factors in
the operations below.

$\bullet$
The space $\Omega_*(\bar M) = \oplus_{p=0}^\infty \Omega_p(\bar M)$
of {\it forms} of various ranks and its further extension
$\Omega^*_*(\bar M) = \oplus_{k=0}^\infty \Omega_*^{\otimes k}$.
This $\Omega_*^*$ plays the same role for  $\Omega_*$
as the Fock space plays for a Hilbert space in a
second-quantized quantum theory
(of course, this is more than a simple analogy).

$\bullet$
Extension of poly-linear operations $\Omega^{\times p}_* \rightarrow \Omega_*$
to linear operators on the Fock space $\Omega^*_* \rightarrow \Omega^*_*$ with the help of {\it lifting formulas}.

$\bullet$
Explicit definition of the elementary operations:
exterior derivative $d$, its conjugate $\partial = d^\dagger$
and exterior product $\wedge$ on $\Omega_*$ and
-- by lifting rule -- on $\Omega_*^*$.
These operations satisfy the number of nilpotency relations:
\be
d^{\circ 2} = 0, \nn \\
\partial ^{\circ 2} = 0, \nn \\
d\circ \wedge + \wedge \circ d = 0
\label{elenil}
\ee
-- the last one is known as Leibnitz rule, and the plus sign
in it is provided by the special adjustment of sign factors
in the definition of lifting.
Two of these elementary operations play the role of the
lowest components of the $A_\infty$ structure,
$m^{(1)} = d$ and $m^{(2)} = \wedge$,
with (\ref{elenil}) realizing the first two relations
(\ref{A2comp}) with $k=2$ and $k=3$, while $\partial$
is used in the construction of the $K$-operator.

$\bullet$
$K$-operator \cite{Kop} satisfies
\be
dK+Kd = I
\ee
where $I$ is the unit operator
on the complement to the kernel of $d$.
It can be explicitly realized as
\be
K(\alpha) = \frac{1}{\Delta(\alpha)}\circ \alpha
\ee
for any linear operator $\alpha$ with invertible
\be
\Delta(\alpha) = \alpha \circ d + d\circ\alpha
\ee
A natural choice for the role of $\alpha$ is
$\alpha = \partial = d^\dagger$.

$\bullet$
Generally speaking, $K(\partial)$ is a non-local
operator, worse than that, it depends in a sophisticated
(and nearly uncontrollable) way on the structure of
simplicial complex.
DMS-construction \cite{DMS} introduces the notion of
{\it strictly-local} operations and the localization
procedure (denoted by square brackets)
for {\it a priori} non-local operations.
Application of this procedure to $\partial$ and
$K(\partial)$, involving the study of the corresponding
strictly-local Laplacian $\left[\Delta([\partial])\right]$
and explicit evaluation of its eigenvectors and
(integer-valued!) spectrum,
provides an explicit construction of a strictly-local
$K$-operator:
\be
K = \frac{1}{\left[\Delta([\partial])\right]}\circ[\partial].
\ee
{\bf This is the central and most non-trivial part of the
DMS construction.}

$\bullet$
The other $k$-linear operations of the $A_\infty$ structure
(\ref{A2}) are made out of $\wedge$ and $K$:
\be
m^{(i)} = (-1)^i \wedge (\wedge \circ K)^{\circ (i-2)} & i \geq 2
\ee
so that entire
\be
D = \bigoplus_{i=1}^\infty m^{(i)} =
U\circ d\circ U^{\circ -1}
\ee
can be considered as a conjugation of the {\it bare}
nilpotent operator $d$ with an operator
\be
U = I + \wedge\circ K
\label{Uop}
\ee
and this conjugation is induced (in a $d$-dependent way,
controlled by the $K$-operator) by exterior product
$\wedge$, satisfying consistency relation (\ref{elenil}).

\section{$A^n_\infty$ structure on simplicial complex
\label{Anstr}}

Generalization to higher nilpotents is straightforward,
and the analogues of eqs.(\ref{elenil})-((\ref{Uop}) are:

\bigskip

$$
\begin{array}{|c|c|}
\hline
&\\
d^3 = 0 & d^n = 0 \\
&\\
d^2\wedge + d\wedge d + \wedge d^2 = 0 &
\ \sum_{k=0}^{n-1}\ d^{n-1-k}\circ\wedge\circ d^k = 0  \\
&\\
\hline
&\\
d^2K + dKd + Kd^2 = I &
\ \sum_{k=0}^{n-1}\ d^{n-1-k}\circ K \circ d^k = I   \\
\multicolumn{2}{|c|}{}\\
\multicolumn{2}{|c|}{
K(\alpha) = \frac{1}{\Delta(\alpha)}\ \alpha}\\
\multicolumn{2}{|c|}{}\\
\Delta(\alpha) = d^2\alpha + d\alpha d + \alpha d^2 &
\ \Delta(\alpha) =
\sum_{k=0}^{n-1}\ d^{n-1-k}\circ\alpha\circ d^k   \\
&\\
\hline
&\\
\alpha = \partial \equiv (d^2)^\dagger &
\alpha = \partial \equiv (d^{n-1})^\dagger\\
\multicolumn{2}{|c|}{}\\
\multicolumn{2}{|c|}{{\rm super-localized}\ [\Delta(\partial)]\
{\rm and\ its\ inverse}}  \\
\multicolumn{2}{|c|}{}\\
\hline
\multicolumn{2}{|c|}{}\\
\multicolumn{2}{|c|}{D = UdU^{-1}} \\
\multicolumn{2}{|c|}{}\\
\multicolumn{2}{|c|}{U = 1 + \epsilon} \\
\multicolumn{2}{|c|}{}\\
\hline
& \\
\epsilon = d \wedge K + \wedge d K + \wedge K d &
\epsilon = \sum\limits_{i,j,k} d ^ i \cdot \wedge \cdot d ^ j \cdot K \cdot d ^ k, \ \ \ i + j + k = n - 2 \\
& \\
\hline
\end{array}
$$

\bigskip

The left column represents the first non-trivial case
of $n=3$, a little more sophisticated formulas for
generic $n$ are written in the right column.
Subscripts $n$ and most of composition signs $\circ$ are
omitted to avoid overloading of formulas.

Denote $\beta_i = \omega_n ^ i$, see \cite{DMS} for details about the other notations.

\paragraph{Higher discrete differential}
$d_n:\ \ \Omega_* \rightarrow \Omega_*$
is defined as follows:
\be
d_n (\sigma) = \sum\limits_{x \notin \sigma}
\beta_{ord(x \rightarrow \sigma \cup x)} \cdot \cup(\sigma, x)
\ee
$\sigma$ is a simplex in $M$, $x$ -- a vertex in $M$. The higher co-diffential is defined as $\partial_n = (d_n^{n-1})^\dagger$.

Of course, some elements of this construction need further detalization. The most important are: higher wedge product, explicit definition of the lifting, inversion of the super-localized Laplace operator $[\Delta]_n$, and construction of higher $K_n$-operator.

\section{Acknowledgements}

Our work is partly supported by Federal Nuclear Energy Agency,
by the joint grant 06-01-92059-CE,  by NWO project
047.011.2004.026, by INTAS grant 05-1000008-7865,
by ANR-05-BLAN-0029-01 project (A.M.), by RFBR grants 07-02-00642 (A.M. and Sh.Sh) and 07-02-01161 (V.D.), by the Grant of Support for the Scientific Schools NSh-8004.2006.2.

\end{document}